# Sketch-to-Layout: A Human-Centric Computational Agent for Constraint-Aware Synthesis of Modular Photobioreactors

*Xiujin Liu[1], ShuqiLi[1], Yuxin Lin[2]*

*[1]Qrafty Technology Inc. [2]University of Michigan*

**Abstract**

Building-integrated photobioreactors (PBRs) offer a promising trajectory for carbon-neutral architecture, yet their deployment is constrained by high-dimensional configuration spaces and complex biological maintenance. This paper presents a modular, human centric PBR facade system powered by a computational framework that reconciles high-level design intent with low-level physical validity. We introduce 'carbon-neutralization bricks' featuring an integrated vessel-and-conduit geometry, where monolithic fluid channels enable 'plug-and-play' assembly. To navigate the combinatorial complexity of 14 distinct modular geometries, we develop a Computational Sketch-to-Layout Agent that formulates layout synthesis as a Constraint Satisfaction Problem (CSP).

Utilizing the CP-SAT engine, the agent treats sparse, intuitive user sketches as soft priors while strictly enforcing hard physical constraints, such as precise port alignment and global topological connectivity. This offloads the structural burden to the agent, allowing non-experts to synthesize valid, fabrication-ready configurations in near real-time. Furthermore, to facilitate autonomous maintenance, we propose a weakly supervised algae health monitoring pipeline. By employing a hybrid CNN-attention backbone and a temporal ranking loss, the system quantifies biological vitality from ordinary photographs without requiring absolute ground-truth labels.

Comprehensive experiments demonstrate that the CSP solver achieves a 95.5% success rate on grid scales up to 15 x 15. Qualitative evaluations confirm the framework's ability to preserve design semantics while ensuring operational integrity. Long-term deployment tests indicate that the vision module produces health trajectories accurately aligned with 14-day biological cycles, suggesting that integrating interactive synthesis with low-cost computer vision can democratize scalable carbon capture systems.

## 1. Introduction

The building sector is a major source of carbon emissions, motivating technologies that turn building surfaces into sites of active carbon capture. Microalgae photobioreactors (PBRs) combine carbon fixation with shading, thermal regulation, and biomass. While early demonstrators such as the BIQ House production (Wurm and Pauli 2016) in Hamburg have proven the functional potential of these systems, the architectural industry has yet to see widespread adoption due to the transition from bespoke engineering to scalable, user-deployed installations.

Recent work (Liu 2025) introduced the concept of "Neutralization Bricks", modular PBR units designed for sustainable architecture that integrate filtration and carbon capture within a prefabricated form factor. While those efforts established the feasibility of the modular hardware, and brought up user customization concept, they revealed a critical "complexity gap" in system-level deployment. Specifically, as the assembly scales from individual units to complex facade grids, two practical difficulties arise that current research has not resolved.

The first challenge is the topological challenge of fluidic connectivity. A functioning PBR assembly requires an airtight pipe network where every module connects to a single air inlet, all neighboring ports are aligned, and there are no boundary leaks. With several "Bricks" prototypes, as the grid size increases, the combinatorial complexity of satisfying these simultaneous conditions, grows exponentially. For a non-expert user, while customization, configuring such a network by hand is not only tedious but prone to single-point failures that render the entire system non-functional. This overwhelming focus on low-level troubleshooting effectively stifles the user’s creative agency; instead of exploring diverse aesthetic and spatial configurations, the user becomes bogged down in mechanical drudgery. This lack of automated assistance strips away the intrinsic joy of DIY assembly, transforming a potential 'play-and-create' experience into a frustrating search for functional validity. Current architectural design tools often lack the semantic understanding of fluidic constraints required to bridge the gap between a non-expert user’s artistic vision and a physically valid layout.

The second one is the maintenance bottleneck in distributed systems. Once installed, the biological health of the algae culture must be monitored over time to ensure optimal carbon capture performance. However, conventional monitoring instruments, such as spectral analyzers or biochemical assays, are costly and impractical for distributed, domestic-scale installations maintained by non-specialists. Without a low-cost, automated feedback mechanism, these "living facades" risk biological decay and operational abandonment.

This paper addresses both challenges by introducing a human-centric computational agent that offloads physical complexity while preserving user agency. We present two primary technical contributions:

- **Sketch-to-Layout Synthesis for User Customization:** We formulate the layout of the pipe network as a Constraint Satisfaction Problem (CSP). By framing user sketches as soft preferences and physical requirements as hard constraints, the system enables users to generate valid, fabrication-ready configurations from rough drawings in near real-time.
- **Weakly Supervised Temporal Monitoring:** We propose a vision-based pipeline that estimates algae health using only temporally ordered photographs. By leveraging the biological prior of monotonic decay through a temporal ranking loss, the model provides actionable health feedback without requiring expensive laboratory labels or expert intervention.

- **From Installation to Stewardship: A User-Friendly Lifecycle Framework:** Beyond technical validation, our framework prioritizes the user experience of biological maintenance, transforming the passive role of an "occupant" into an active "steward." By making the complexities of carbon capture "visible" and "manageable" for non-experts, this research significantly bolsters the potential for the large-scale, democratic adoption of living architectural skins.

## 2. Related Works

### 2.1 Photobioreactor (PBR) Building Components

Building-integrated algae cultivation has progressed from early demonstrators such as the BIQ House in Hamburg (Wurm and Pauli 2016) through a variety of facade-integrated PBR typologies, including pipe-shaped systems for biofuel and well-being (Kim 2013) and the Pleura Pod air-filtration concept (Lee 2014). While the thermal, optical, and carbon-capture benefits of these systems are increasingly well documented, scalability and maintenance remain open challenges. Liu advanced a modular approach based on prefabricated "neutralization bricks" (Liu 2025), demonstrating modular-system feasibility but leaving unaddressed the system-level problem of configuring many modules into a pipe network that is globally connected, correctly aligned, and airtight.

### 2.2 Constraints Based computational Design

The constraint satisfaction problem (CSP) has a long history in architectural layout, from early floor-plan generators subject to adjacency and area rules (Hoffmann 2005) to more recent residential-complex solvers incorporating daylight and privacy evaluation (Sherkat et al. 2023). Extensions to heterogeneous domains of physics-and-kinematics (Mirzendehdel et al. 2019) and generative layout systems (Li et al. 2023) have further broadened the scope. Thus, it shows promise for applying CSP to complex network configurations in modular building systems. Systems use parametric and constraint-based models to automatically generate diverse building layouts that still satisfy practical design requirements. (Li et al. 2023)

### 2.3 Computer Vision for Microalgae Monitoring

Deep learning has been applied to algae monitoring by species detection from microscopic images (Abdullah et al. 2019) dedicated marine-microalgae datasets (Zhou et al. 2023), and low-cost cell-concentration estimation through classical image processing (Borisova et al. 2025). However, most microalgae monitoring methods are designed for controlled laboratory conditions and require specialized tools and expertise, limiting the large-scale deployment of PBR building systems.

### 2.4 Gaps

Significant gaps remain in the development of modular photobioreactor (PBR) systems, where scaling to building or city-level applications is fundamentally hindered by the combinatorial complexity of airtight fluidic interconnections. While modular "neutralization bricks" have established a physical foundation for carbon-neutral facades, the lack of intuitive design integration restricts user-driven flexibility and participatory design in real-world settings. Conventionally, the rigorous technical requirements for assembly impose a steep barrier to entry that necessitates specialized expertise and discourages distributed adoption. Furthermore, the operational reliance on laboratory-grade monitoring tools rather than low-cost, automated feedback loops complicates long-term maintenance, creating an urgent need for computational frameworks that can offload topological complexity to interactive agents. By bridging the gap between creative architectural intent and rigorous physical validity, these gaps can be addressed to facilitate the democratized deployment of resilient and adaptive PBR systems.

## 3. Methodology

### 3.1 System Overview

The proposed modular PBR system is conceptualized as a multi-layered computational agent that mediates between high-level architectural intent and low-level biological performance. Building upon the physical foundation established by Liu (2025), established the efficacy of individual "Neutralization Bricks" as prefabricated units for carbon sequestration, this framework shifts the focus toward a systemic, multi-agent assembly capable of large-scale facade integration. The transition from a single unit to a functional collective necessitates a hierarchical architecture that simultaneously manages physical fluidics and digital design preferences. The architecture of the system is organized into three primary operational modules: The Global Grid Module, The Sketch-to-Layout Module, The Maintenance Module.

### 3.2 Global Grid Module

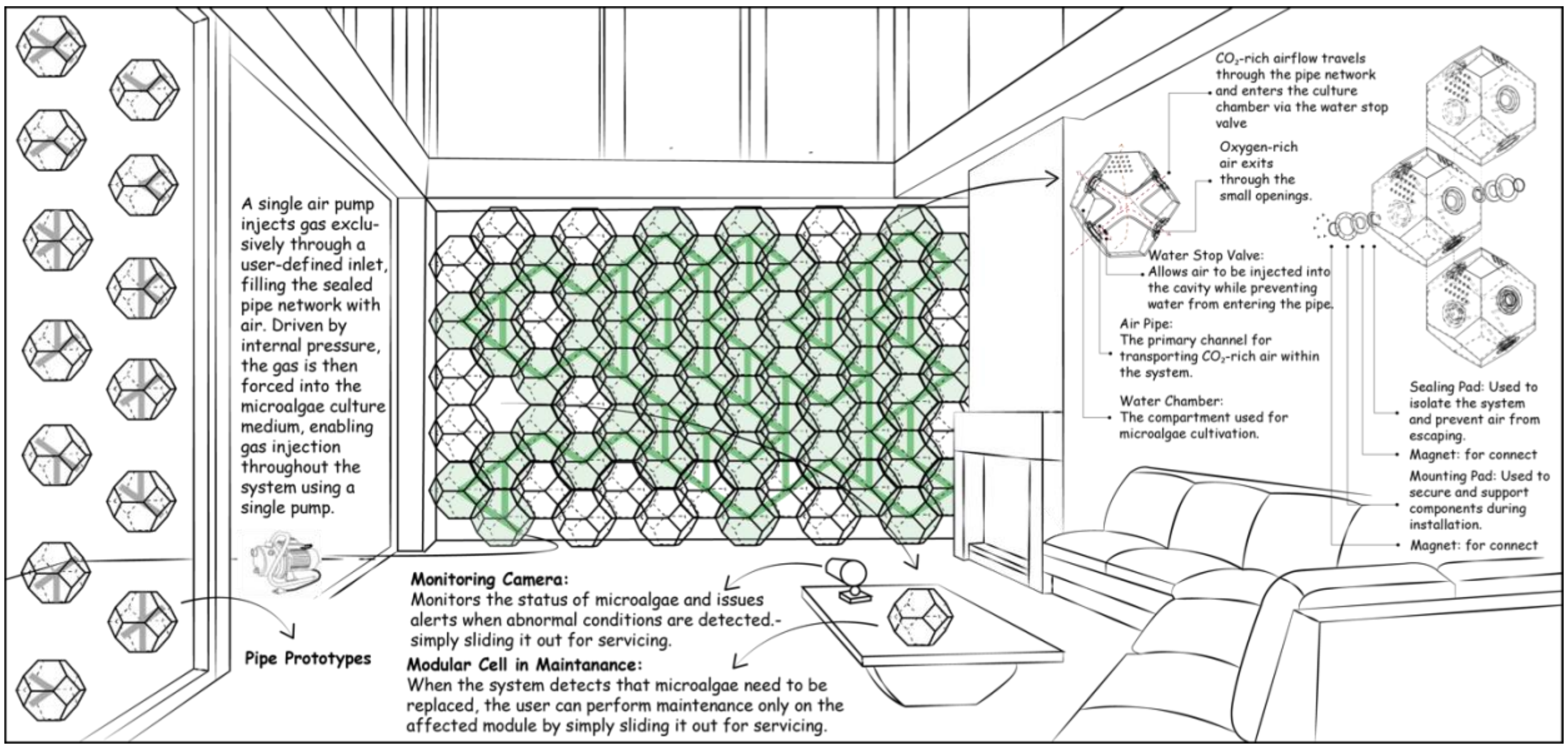

**Figure 1. Architecture of the modular PBR facade system, showing the module grid, pipe connectivity, air circulation path, and monitoring integration.**

The Global Grid Layer establishes the foundational logic of the assembly by defining spatial boundaries and air injection points for the entire fluidic network. This layer treats the facade as a discretized hexagonal grid where "carbon-neutralization bricks" utilize an integrated vessel-and-conduit design with fluid channels embedded directly into the monolithic module body. To maximize creative freedom and user engagement, the system offers a library of 14 distinct modular geometries that allow for "plug-and-play" assembly and disassembly via magnetic support structures, enabling non-experts to intuitively configure complex facades without specialized tools. As illustrated in Figure 1, the entire system is engineered to function as a single, airtight pipe network rooted at a user-designated inlet; driven by internal pressure, gas is forced exclusively through these sealed conduits into the microalgae culture medium. This global integrity ensures that even large-scale assemblies remain physically valid, while the modular logic permits targeted maintenance where users can simply slide out an individual cell for servicing without disrupting the rest of the grid.

## 3.3 The Sketch-to-Layout Module

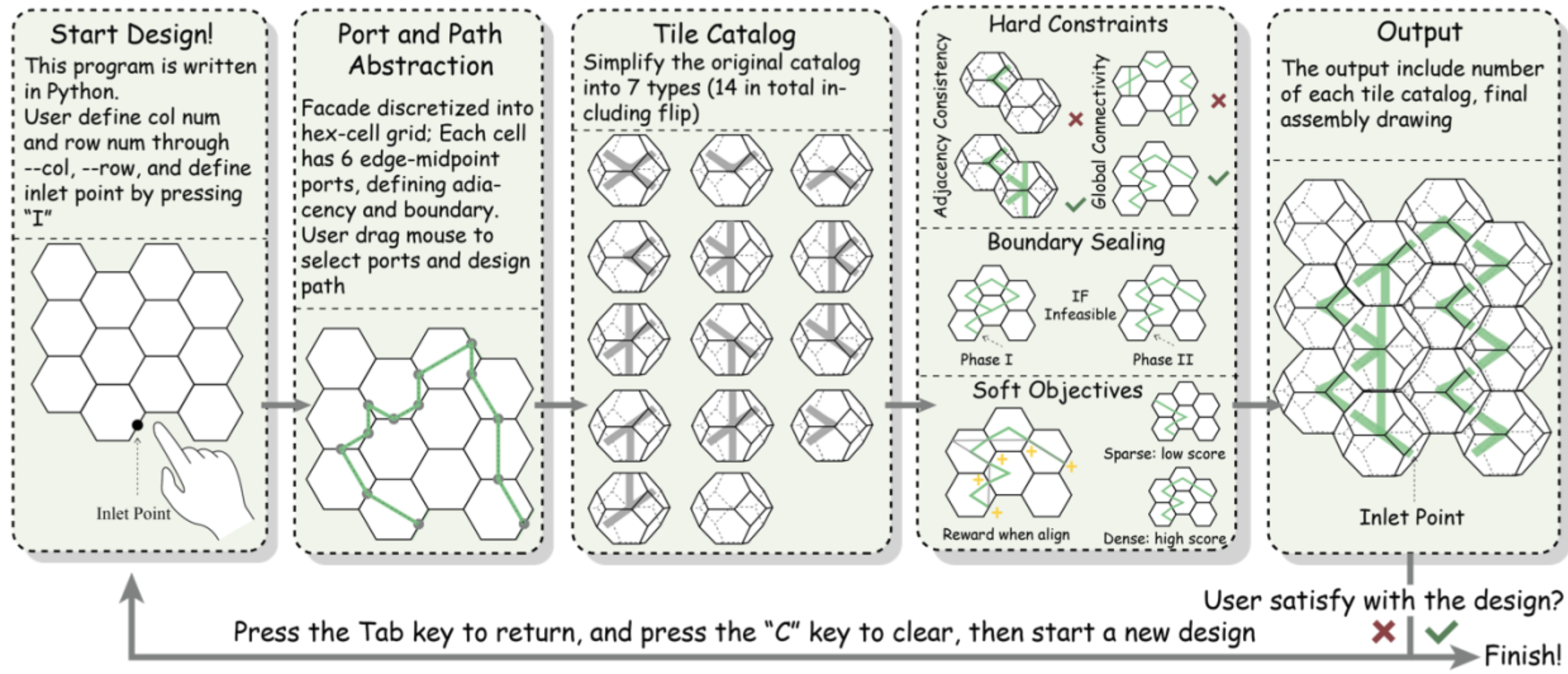

**Figure 2. Sketch-to-Layout Module framework.**

The Sketch-to-Layout Module acts as the fluidic engine of the system, managing the transition from abstract design to functional infrastructure. It serves as a vital bridge to lower the barrier between non-expert users and domain-specific professional knowledge. This Module's primary role is to synthesize "Pipe Layout Prototypes", the internal routing configurations that ensure a continuous, airtight flow path across dozens or hundreds of modules. By allowing users to express design intent intuitively, the system shifts the burden of technical specification from the human to a computational agent. To be more specific, a valid pipe network in a modular PBR facade must satisfy three conditions at once: adjacent modules must open matching ports on their shared edge, the outer boundary must be sealed except at the air inlet, and every module must be reachable from that inlet through a continuous connected path. The combinatorial complexity of satisfying physical constraints grows exponentially with the number of modules, far exceeding what a non-expert can manage by hand. Conventionally, these rigorous technical requirements impose a steep barrier to entry, significantly increasing the difficulty for users attempting custom designs or autonomous installation. To resolve this, the system employs an interactive Sketch-to-Layout workflow that empowers non-expert users to participate in complex system design without requiring specialized engineering knowledge. Instead of requiring the manual construction of airtight networks, our system allows users to indicate preferred connectivity tendencies through free-form sketches. These inputs are then mediated by a framework that enforces rigorous physical and fabrication constraints, ensuring airtightness and global connectivity while preserving the aesthetic spirit of the user's original intent.

The overall framework is illustrated in Figure 2, which details the transformation of sparse user inputs into feasible, fabrication-ready configurations. To guarantee that the generated geometry forms a physically valid pneumatic chamber, we formulate pipe generation as a Constraint Satisfaction Problem (CSP). Unlike heuristic methods that often produce isolated or non-functional fragments, our approach enforces system-level validity through three rigorous Hard Constraints: port alignment (A pipe connection across a shared module edge is valid only if both modules open the corresponding ports. One-sided openings are rejected), boundary sealing (Every port on the outer perimeter of the grid defaults to closed. The sole exception is the user-designated gas inlet, preserving the pressure differential needed for circulation), and global connectivity (A flow-conservation formulation ensures that all active modules belong to a single connected component rooted at the inlet. Isolated loops and orphan branches are prohibited).

Algorithm 1 formalizes the synthesis logic. Subject to these constraints, the system reads user sketches not as literal specifications, but as a set of "soft" design preferences to be honored wherever physically possible. Specifically, the solver maximizes an objective that rewards agreement with the user's sketch and the overall pipe coverage. Because these physical constraints are strictly enforced, the output is guaranteed to be valid irrespective of the initial sketch quality. To ensure robustness, the solver first seeks a perfectly sealed solution; if none is found within the time budget, it re-solves using a relaxed boundary constraint that permits controlled vents at selected perimeter ports. This fallback mechanism ensures that the system returns a usable result in nearly all cases. By offloading the burden of physical validity to a computational agent, the system significantly increases user engagement and makes the design process "user-friendly," allowing users to iterate rapidly through the interactive loop described in Algorithm 2.

The proposed strategy is not limited to microalgae PBRs. The framework's topological logic is readily adaptable to other fluid-based facade systems—such as HVAC or hydraulic active envelopes and can even be recontextualized to optimize human movement streamlines, demonstrating robust versatility across both mechanical and spatial domains.

**Algorithm 1:** Adaptive Pipe Synthesis Logic

**Data:** Grid $\mathcal{C}$, Inlet $P_{in}$, Sketch $\mathcal{S}$
1 **Function** `Synthesize(`*AllowVents*`):`
2     **Input:** $\mathcal{L}$;
3     Init Solver $\mathcal{M}$ over grid $\mathcal{C}$;
4     **Subject to:**;
5     1. *Topology*: Adjacent tiles must connect seamlessly (Open-to-Open);
6     2. *Continuity*: All pipes must flow from $P_{in}$ (No isolated loops);
7     3. *Boundary*: Seal all edges; allow optional vents if $\mathcal{L}$ is True;
8     **Maximize:** Alignment with $\mathcal{S}$ + System Volume;
9     **return** $\mathcal{M}$.Solve();

10 **Adaptive Two-Pass Generation:**;
11 $\mathcal{L} \leftarrow$ False ; `// Attempt 1: Perfectly Airtight`
12 **if** *AllowVents* = *Null* **then**
13     $\mathcal{L} \leftarrow$ True ; `// Attempt 2: Relaxed Fallback`
14 **return** $\mathcal{L}$;

**Algorithm 1. Adaptive Pipe Synthesis Logic**

**Algorithm 2:** Sketch-to-Fabrication Interaction Loop

```
   Data: Grid Resolution R, Tile Constraints C
   Result: Fabrication-Ready Geometry
1  Initialize Grid G and Port Set P_all;
   // Real-time Interaction Loop
2  while Session Active do
3      if User Drags Mouse (Sketching) then
4          Get cursor pos (mx, my);
5          g_nearest ← NearestNeighbor(mx, my, P_all);
6          Add g_nearest to Sketch Set S and update canvas;
7      else if User Selects Inlet then
8          Validate boundary constraint and set Inlet I;
9      else if User Triggers Synthesis then
           // Call Algorithm 1 for layout generation
10         L ← AdaptiveSynthesis(G, S, I);
11         if L is Valid then
12             Instantiate 3D geometric tiles from L;
13             Output: Watertight model for manufacturing;
14         else
15             Display "Infeasible" warning and prompt for edits;
```

**Algorithm 2. Sketch-to-Fabrication Interaction Loop.**

### 3.4 The Maintenance Module

Precise algae health indices typically require professional-grade instrumentation such as spectral analyzers, chlorophyll concentration meters, or biochemical sampling tools, which often demand expert calibration and interpretation. In contrast, our goal is to provide the simplest and lowest cost mechanism for end users to assess algae conditions in real modular PBR façades. Therefore, to estimate algae conditions without precise health annotations, we propose a weakly supervised framework. This approach uses only RGB images and temporal ordering to learn a latent embedding of brick-level algae health, enabling relative comparisons between temporally sequenced images.

Our goal here is to give building occupants a tool that turns an ordinary photograph into a meaningful health estimate of the microalgae, requiring nothing beyond a camera and a date. For a sealed PBR module, in principle, the vitality of algae decreases monotonically as nutrients are consumed and metabolic waste accumulates. Therefore, an image of algae captured on a later day should receive a lower health score than an image captured on an earlier day, even when the absolute health value on either day is unknown. This temporal prior pro- vides a supervisory signal without requiring expert labeling. Figure 3 outlines the pipeline, which proceeds in four stages.

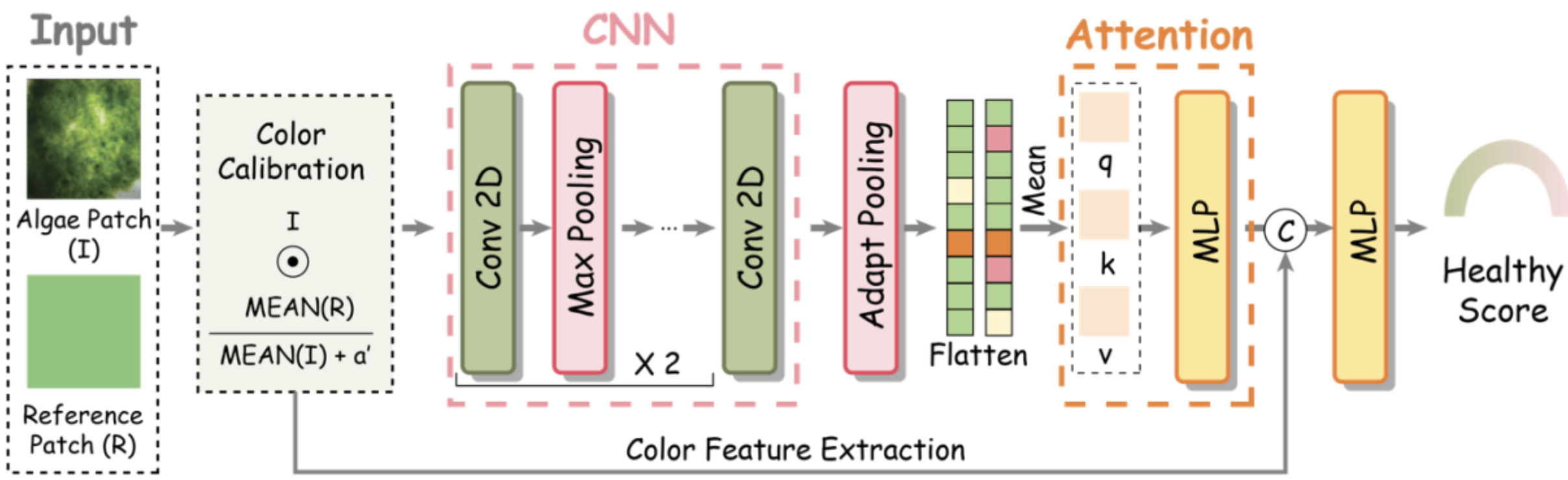

**Figure 3. Weakly supervised algae monitoring pipeline. Input images undergo color calibration against a reference patch, feature extraction through a hybrid CNN-attention backbone and explicit color descriptors, and temporal ranking to yield a continuous health score.**

**Color Calibration:** Since outdoor illumination fluctuates with weather conditions and time of day, raw pixel values are unreliable indicators of biological status. To address this, each image is normalized against a reference patch of a known, fixed color embedded directly into the module frame. As shown in Equation 1., by comparing the captured reference patch to its standard values, the system applies a calibration factor to the entire image. This ensures that whether a photo is taken under a sunset or a cloudy sky, the algae color is restored to a consistent baseline for fair analysis.

$$\tilde{I} = I \odot \left(\mathrm{MEAN}(R)\right)/(\mathrm{MEAN}(I) + a') \quad (1)$$

**Feature Extraction:** The calibrated image is processed through a "hybrid brain" pairing Convolutional Layers with an Attention mechanism. While the convolutional layers capture local textures—such as algae clumping or adhesion to the vessel walls—the attention mechanism aggregates the global context of the panel. Alongside these learned features, we compute explicit color descriptors to monitor ratios between red, green, and blue channels.

**Health Score Prediction and Temporal Ranking Loss:** The system maps fused textural and color features into a continuous health score $p$, providing an intuitive indicator of biological vitality for the user. To train this model without costly laboratory labels, we implement a temporal ranking loss based on the biological prior of monotonic decay; in a sealed environment, algae health inevitably decreases as nutrients are consumed. By comparing pairs of images *(i, j)* from the same module where $t_j > t_i$, the loss penalizes any configuration where the later image receives a higher score than the earlier one:

$$L = 1/|P| \sum \left(i < j, t_j > t_i\right) \max\left(0, 1 - \left(p_j - p_i\right)\right)$$

Through this self-correcting logic, the network learns to accurately track degradation trajectories using only a camera and a date stamp, facilitating long-term maintenance without expert intervention. In deployment, a user photographs facade modules at whatever intervals are convenient, attaching only a date. After an initial collection phase, the model scores new images on the fly, flagging modules whose health has dropped below a maintenance threshold.

## 4. Experiments

In this section, we evaluate the two core components: The Sketch-to-Layout Module and The Maintenance Module. The practice case study is shown in Figure 4. (Liu 2025)

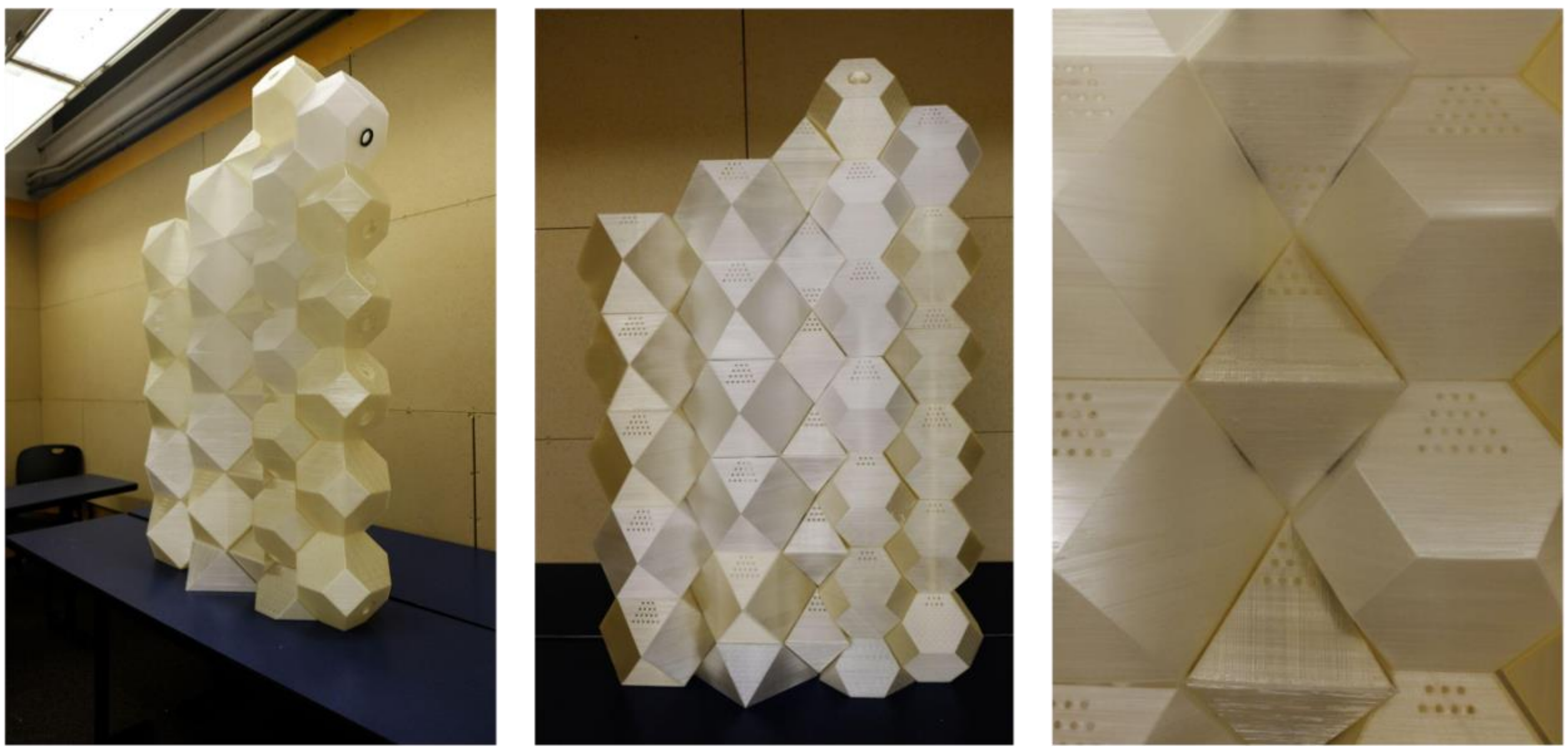

**Figure 4. Physical case study of the modular PBR facade assembly.**

### 4.1 The Sketch-to-Layout Module

**Setup:** A dataset of 500 synthetic user sketches was constructed to cover a broad spectrum of design scenarios: simple linear strokes, branching patterns, disconnected fragments, and deliberately contradictory inputs. The solver employs the CP-SAT engine from Google OR-Tools under a 10-second time limit per instance. Grid sizes range from 5 x 5 (component scale) to 15 x 15 (facade scale). All experiments ran on a MacBook Pro with an Apple M2 Pro chip; no GPU was used.

**Metrics:** Three quantities are reported: the success rate (fraction of inputs for which the solver returns a valid, airtight layout), the mean Intersection over Union (IoU) between the sketch and the generated pipe network (a proxy for how faithfully user intent is preserved), and the mean wall-clock time per solve.

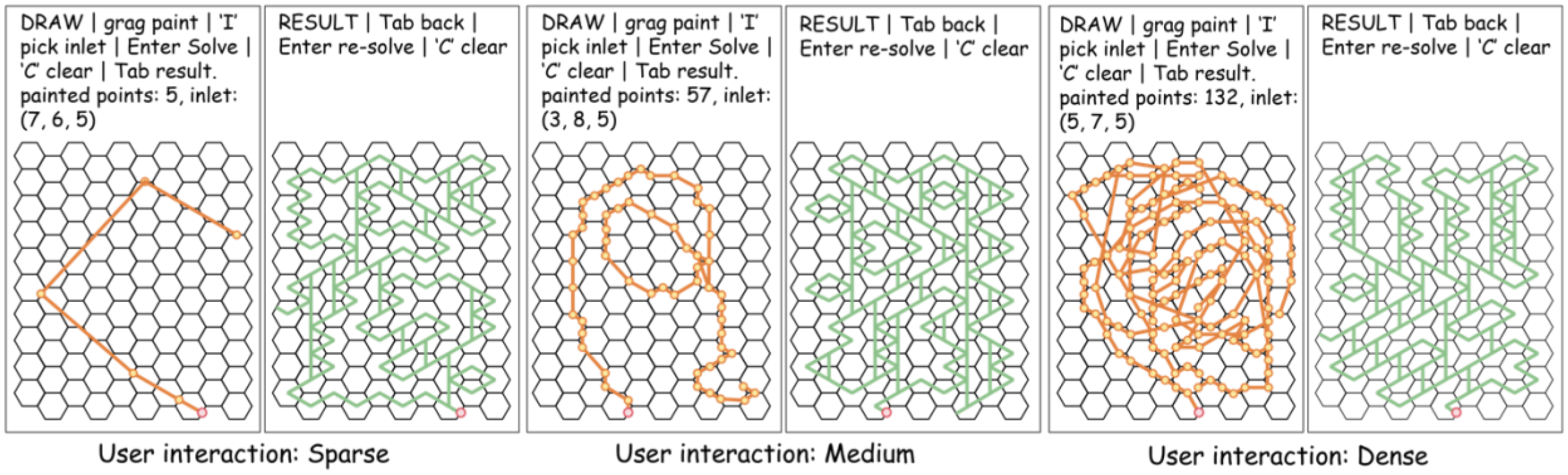

**Figure 5. Pipe layouts generated from increasingly dense user sketches (left to right). Valid, sealed configurations are produced consistently regardless of input complexity**.

**Results:** Figure 5. displays representative outputs across the sparse-to-dense sketch spectrum. Table 1. collects the aggregate statistics. The solver finds a valid configuration for 95.5% of inputs, confirming robustness across diverse and often adversarial sketch conditions. The mean IoU of 0.375 warrants comment: because the solver is obliged to override sketch preferences whenever they conflict with hard constraints, perfect agreement is neither expected nor desirable. What the metric does confirm is that the solver makes use of the sketch wherever doing so is compatible with physical validity. At 3.3 seconds on average, solve times are well within the range acceptable for interactive design iteration.

**Table 1: Aggregate pipe generation results over 500 synthetic sketches.**

| Success Rate (%) | Mean IoU | Mean Runtime (s) |
|---|---|---|
| 95.5 | 0.375 | 3.334 |

4.2 The Maintenance Module

**Dataset:** The monitoring module is evaluated on a set of microalgae images, documenting the entire life cycle over 14 days. The 325 images were captured at irregular intervals across morning, noon, and evening conditions, mirroring the uneven schedules characteristic of real maintenance practice.

**Implementation:** Given the modest dataset size, an aggressive augmentation pipeline-random crops, horizontal and vertical flips, color jitter, and Gaussian blur-was applied to reduce overfitting. All images were resized to 224 x 224 and normalized with ImageNet statistics. Training used the Adam optimizer with a learning rate of $1 \times 10^{-4}$, weight decay of $1 \times 10^{-5}$, and batch size of 64, running on an NVIDIA RTX 3090.

**Evaluation:** Classification accuracy is not the appropriate metric here; what matters is whether the model's scores decline in step with the known biological trajectory. A health score normalized to the interval [0,1] is defined, 1.0 corresponding to a freshly inoculated culture and 0.0 to the end-of-life. The model is applied to all 325 images in chronological order to reconstruct the full health trajectory.

**Results:** The resulting curve, shown in Figure 6. represents a clear monotonic decline over the 14-day window. Inset photographs at selected time points visually corroborate the trend: bright green biomass at the outset gives way to progressively browner, thinner cultures. The correspondence between computed scores and observable degradation confirms that temporal ordering alone provides sufficient supervision for the task. From a practical point of view, the result indicates that building occupants equipped with nothing more than a phone camera can obtain actionable health feedback for individual facade modules.

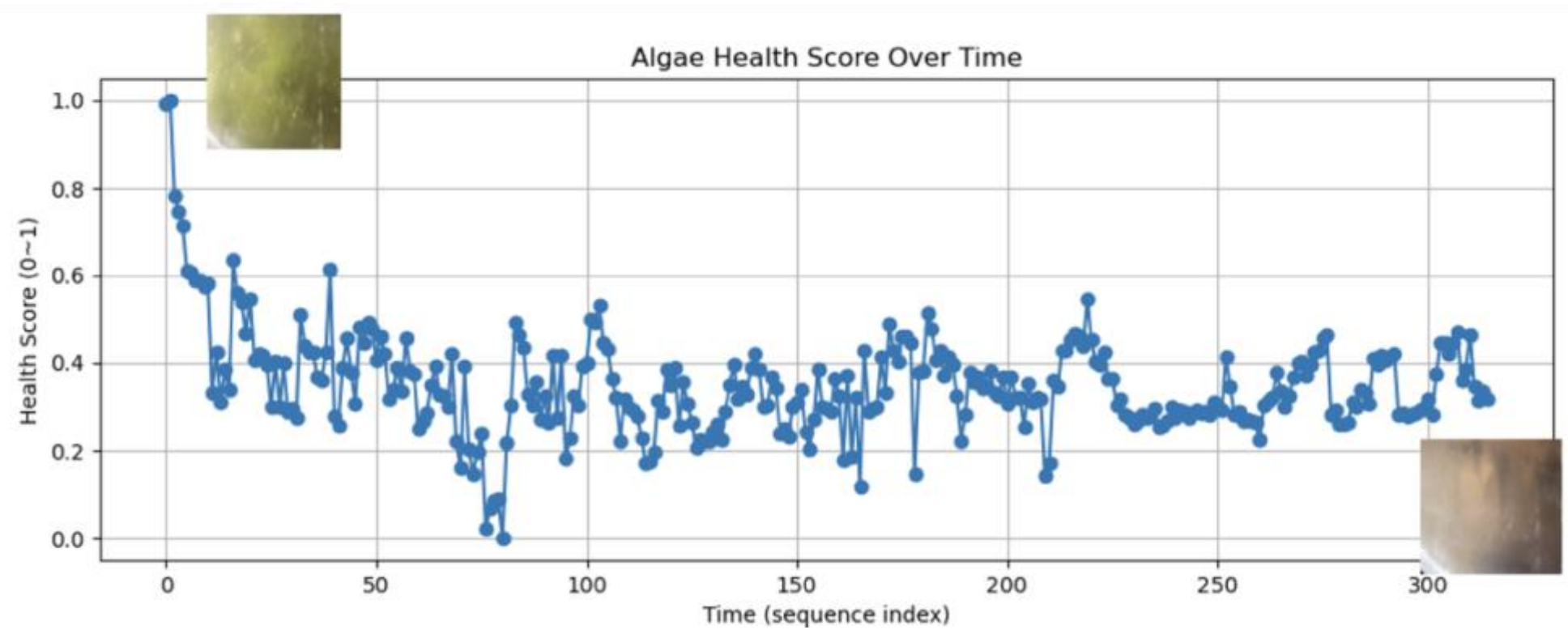

**Figure 6. Health score trajectory over 14 days. The monotonic decline aligns with visible algae degradation, validating temporal ordering as a supervisory signal for health estimation.**

## 5. Conclusion

In general, this paper proposes a user-centered computational pipeline for the deployment and maintenance of microalgae-based modular photobioreactor (PBR) facade systems. Addressing the challenges of scalability, deployability, and maintenance in existing PBR research, the proposed method allows users to configure airtight pipe networks through sketch-based interaction, while ensuring physical validity under fabrication and operational constraints. By interpreting sketches as soft design preferences, and global connectivity and airtightness requirements as hard constraints, the system easily balances user intent with physical laws of the real world. In parallel, a vision-based microalgae monitoring module provides component-level feedback, supporting long-term operation and reducing reliance on expert intervention.

Future work will extend this pipeline toward automated production and maintenance. This includes exploring new printing methods and integrating the design framework with robotic or digital fabrication workflows for automated manufacturing and assembly. In addition, robot-assisted maintenance strategies, such as inspection, cleaning, and module replacement, will be investigated using real-time microalgae monitoring data. These efforts aim to reduce operational complexity and support scalable, resilient, and adaptive PBR façade systems.